\documentclass[conference]{IEEEtran}
\IEEEoverridecommandlockouts

\usepackage{cite}
\usepackage{amsmath,amssymb,amsfonts}
\usepackage{algorithmic}
\usepackage{graphicx}
\usepackage{textcomp}
\usepackage{xcolor}

\usepackage{multirow}
\usepackage{tabularx}
\usepackage{enumitem}
\usepackage{subcaption}
\usepackage{booktabs}  
\usepackage{threeparttable}
\usepackage{colortbl}
\usepackage{siunitx}
\usepackage{tikz}
\usepackage{filecontents}

\def\BibTeX{{\rm B\kern-.05em{\sc i\kern-.025em b}\kern-.08em
    T\kern-.1667em\lower.7ex\hbox{E}\kern-.125emX}}

\begin{document}

\title{Dynamic Pacing for Real-time Satellite Traffic}

\author{\IEEEauthorblockN{Aashish Gottipati\textsuperscript{1} and Lili Qiu\textsuperscript{1}\textsuperscript{2}}
\IEEEauthorblockA{\textsuperscript{1}Department of Computer Science, University of Texas at Austin, Austin, Texas \\ \textsuperscript{2}Microsoft Research Asia, Shanghai, China}}

\newcommand\copyrighttext{%
\footnotesize \textcopyright 2025 IEEE. Personal use of this material is permitted.
Permission from IEEE must be obtained for all other uses, in any current or future
media, including reprinting/republishing this material for advertising or promotional
purposes, creating new collective works, for resale or redistribution to servers or
lists, or reuse of any copyrighted component of this work in other works.}

\newcommand\copyrightnotice{%
\begin{tikzpicture}[remember picture,overlay]
\node[anchor=south,yshift=10pt] at (current page.south)
{\fbox{\parbox{\dimexpr\textwidth-\fboxsep-\fboxrule\relax}{\copyrighttext}}};
\end{tikzpicture}%
}

\maketitle

\begin{abstract}
Google's congestion control (GCC) has become a cornerstone for real-time video and audio communication, yet its performance remains fragile in emerging Low Earth Orbit (LEO) networks. In this paper, we study the behavior of videoconferencing systems in LEO constellations. We observe that video quality degrades due to inherent delays and network instability introduced by the high altitude and rapid movement of LEO satellites, with these effects exacerbated by WebRTC's conventional ``one-size-fits-all'' sender-side pacing queue management. To address these challenges, we introduce a data-driven queue management mechanism that tunes the maximum pacing queue capacity based on predicted handover activity, minimizing latency during no-handover periods and prioritizing stability when entering periods of increased handover activity. Our method yields up to $3$x improvements in video bitrate and reduces freeze rate by $62\%$ in emulation, while delivering up to a $41\%$ reduction in freeze rate and $40\%$ decrease in mean packet loss on real Starlink constellations compared to WebRTC's default pacing queue policy.
\end{abstract}


\section{Introduction}
\label{sec:introduction}

Real-time communication (RTC) applications centered around audio and video have become central to global communication. As reported in~\cite{businessresearchinsights2025webrtc}, over $1.7$ billion individuals utilized WebRTC-based~\cite{webrtc} applications in 2023 alone, with continued growth anticipated due to advancements in generative AI. Simultaneously, Low Earth Orbit (LEO) satellite constellations are emerging as critical infrastructure for ubiquitous, high-speed internet access.

While the convergence of LEO and RTC systems promises seamless global RTC, a fundamental mismatch remains: standard RTC stacks were designed for the stable, predictable conditions of terrestrial networks-- not the constantly shifting dynamics of space. Additionally, LEO constellations are not a monolith and vary significantly between providers, from sparse constellations~\cite{ASTSpaceMobile} comprised of tens to hundreds of large satellites with infrequent handovers (e.g.,\ on the order of minutes) to dense systems such as Starlink that leverage thousands of relatively small satellites and rapid handover activity (e.g.,\ approximately every $15$ seconds~\cite{tanveer2023making}).
Current systems often naively apply terrestrial RTC stacks directly to LEO networks; however, the high altitude, orbital dynamics of LEO satellites, and frequent handover activity introduce burst latency and buffer bloating~\cite{tao2023transmitting}, challenging terrestrial methods and degrading quality of experience (QoE) for end users.

\begin{figure}[t!]
    \centering
    \begin{subfigure}[b]{0.48\columnwidth}
        \centering
        \includegraphics[width=\linewidth, trim={0 0cm 0 0cm}, clip]{"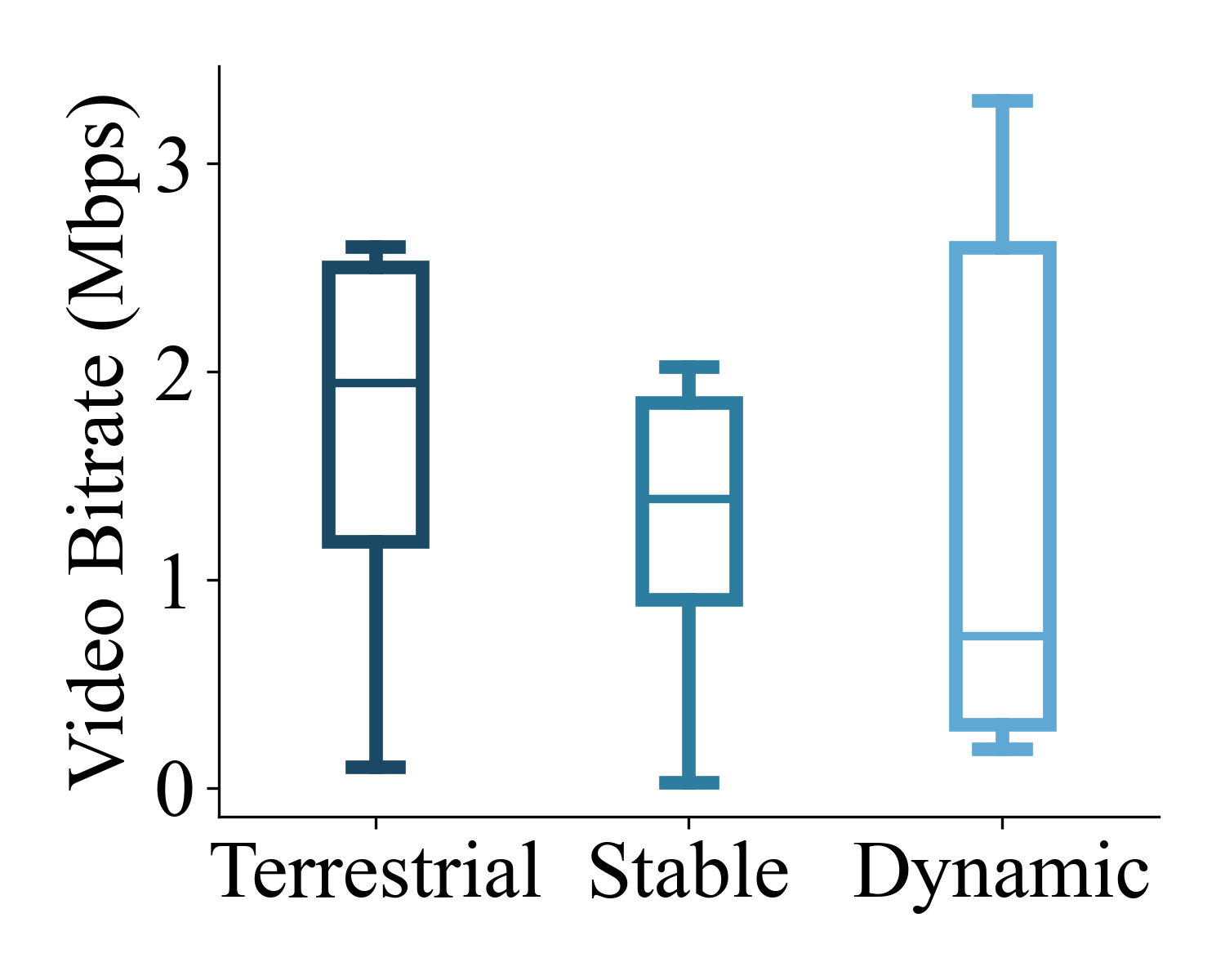"}
        \caption{Video Bitrate.}
        \label{fig:initial-bitrate}
    \end{subfigure}
    \hfill
    \begin{subfigure}[b]{0.48\columnwidth}
        \centering
        \includegraphics[width=\linewidth, trim={0 0cm 0 0cm}, clip]{"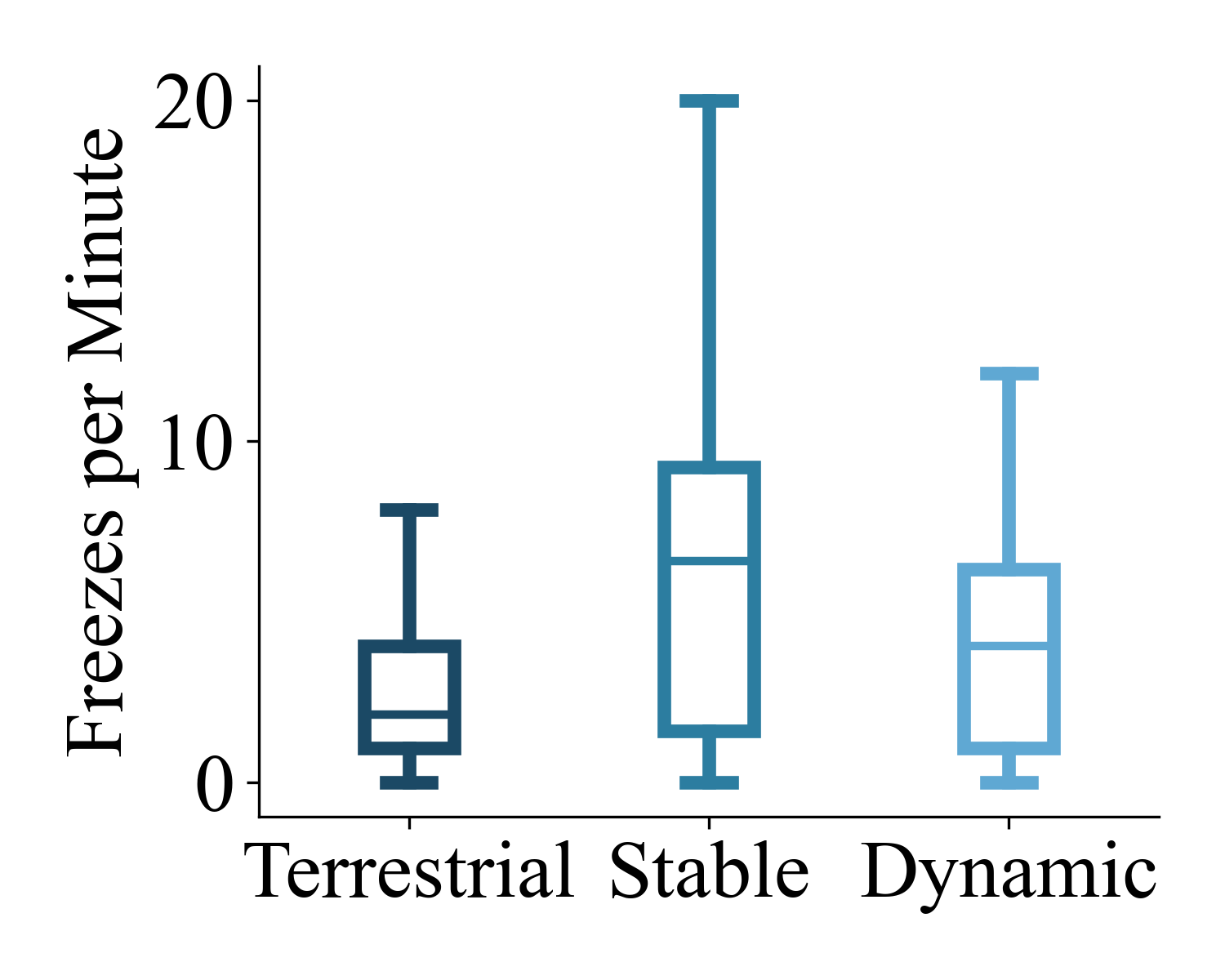"}
        \caption{Video Freeze Rate.}
        \label{fig:initial-freeze}
    \end{subfigure}
    \caption{Two distinct failure modes of WebRTC's default queue policy in LEO networks: self-inflicted delays during stable periods (few handovers) and high instability during dynamic periods (frequent handovers).}
    \vspace{-5mm}
    \label{fig:initial_exploration}
\end{figure}

To illustrate the fragility of RTC systems in LEO environments, we conduct approximately $50$ hours of WebRTC~\cite{webrtc} (an open-source framework for real-time applications) videoconferencing calls across emulated terrestrial mobile links (5G and 4G) and LEO constellations. We leverage Mahimahi~\cite{netravali2015mahimahi} and previously collected mobile network traces to emulate terrestrial links, and utilize StarryNet~\cite{lai2023starrynet}-- a high-fidelity integrated satellite-terrestrial network (ISTN) emulator-- to emulate LEO conditions. We summarize our results in Figure~\ref{fig:initial_exploration}.

Our initial analysis reveals that default WebRTC configurations negatively impact Google's Congestion Control (GCC) algorithm and result in a near $50\%$ increase in freeze rates during stable LEO conditions compared to dynamic scenarios with frequent handovers (see Figure~\ref{fig:initial_exploration}). We identify the packet pacer's static pacing policy as a critical, yet overlooked, contributor to this regression. While the sending queue holds encoded frames, the pacing queue times the release of the client's processed packets onto the network. Originally designed for low-latency terrestrial networks, WebRTC's default pacing policy acts too conservatively during stable LEO periods. This cautious pacing behavior introduces self-inflicted queuing delays at the sender, causing packets for critical video frames to miss their deadlines at the receiver. Conversely, during dynamic periods with frequent handovers, the same policy becomes overly aggressive, causing excessive packet loss. This failure mode demonstrates that the core challenge is not parameter tuning, but the inability of static pacing policies to adapt to the dynamic environment of LEO networks. Our work focuses on managing the pacing queue, which for the remainder of this paper we refer to as queue management.

This paper argues that a ``one-size-fits-all'' queue policy is fundamentally misaligned with the time-varying structure of LEO networks. We frame sender-side queue management as a dynamic decision problem where the optimal strategy adapts to the connection's predictable orbital context. Our key insight is that while pacing decisions cannot alter the movement of LEO satellites, they must be optimized for it. To achieve this, we design a novel \emph{handover-aware} queue management mechanism using offline imitation learning (IL). This approach allows us to train a lightweight and interpretable policy from real-world network logs, making it suitable for direct integration into the WebRTC stack. Specifically, we construct an expert policy by clustering logged decision intervals based on observed handover activity and labeling each cluster with queue sizes based on empirical QoE. Our policy network, a Transformer-based encoder, is trained via supervised learning to map network context to the expert's chosen queue size. The policy's objective is twofold: minimize latency during stable periods and prioritize stability during frequent handovers.

Our formulation has several key benefits: (1) it avoids the complexity of non-stationary value estimation and counter-factual reasoning-- as is common in full offline reinforcement learning (RL); (2) it naturally restricts learning to expert-demonstrated behavior, improving interpretability and reliability; (3) by executing pacing decisions in near real-time, it may utilize robust modeling tools-- such as the SGP4 orbital propagation model and Two-line element set (TLE) data-- to inform queuing decisions without impacting the real-time performance of WebRTC.

Our contributions are as follows:
\begin{itemize}[topsep=0pt]
    \item We identify a key failure mode of RTC systems in LEO networks stemming from ``one-size-fits-all'' sender-side queue limits.
    \item We design a \textit{handover-aware} policy, trained via offline IL, that dynamically adapts WebRTC’s pacing queue based on predicted satellite handover patterns. 
    \item We demonstrate that our approach significantly outperforms WebRTC's default
configuration: up to $3$x improvements in video bitrate and a $62\%$ reduction in
freeze rate in emulated sparse LEO conditions and up to a $41\%$ reduction in freeze
rate and $40\%$ reduction in mean packet loss in live dense Starlink deployments.
\item We plan to release a public dataset of $12.5$ hours of LEO network traces and an open-source, reproducible Cloudlab~\cite{duplyakin2019design} profile to foster further research.
\end{itemize}

\section{Related Work}
\label{sec:related}

Low-latency communication over LEO networks is challenging due to frequent handovers and longer end-to-end delays. Prior work has largely focused on adaptive bitrate (ABR) for live streaming.~\cite{zhao2024low} exploits handover patterns to adjust video bitrate and playback speed, while SARA~\cite{fang2024robust} provides handover feedback to ABR algorithms, reducing rebuffering by modulating playback speed. However, these approaches rely on client buffering, which is incompatible with interactive RTC where sub-second latency and immediate reaction to disruptions are required. Within RTC, existing efforts mainly enhance error correction~\cite{chen2022rl, lee2022r} and rate control~\cite{agarwal2025mowgli, fang2019reinforcement, gottipati2023offline, gottipati2024balancing} mechanisms. In contrast, our work is the first to dynamically adapt the sender-side pacing queue based on predicted handover activity. We show that our \textit{handover-aware} approach can significantly improve video quality for interactive applications such as videoconferencing, where playback buffering is unavailable.

\section{Methods}
\label{sec:methods}
\begin{figure}[t!]
    \centering
    \includegraphics[width=0.75\columnwidth, trim={0cm, 0cm, 0cm, 0cm}, clip]{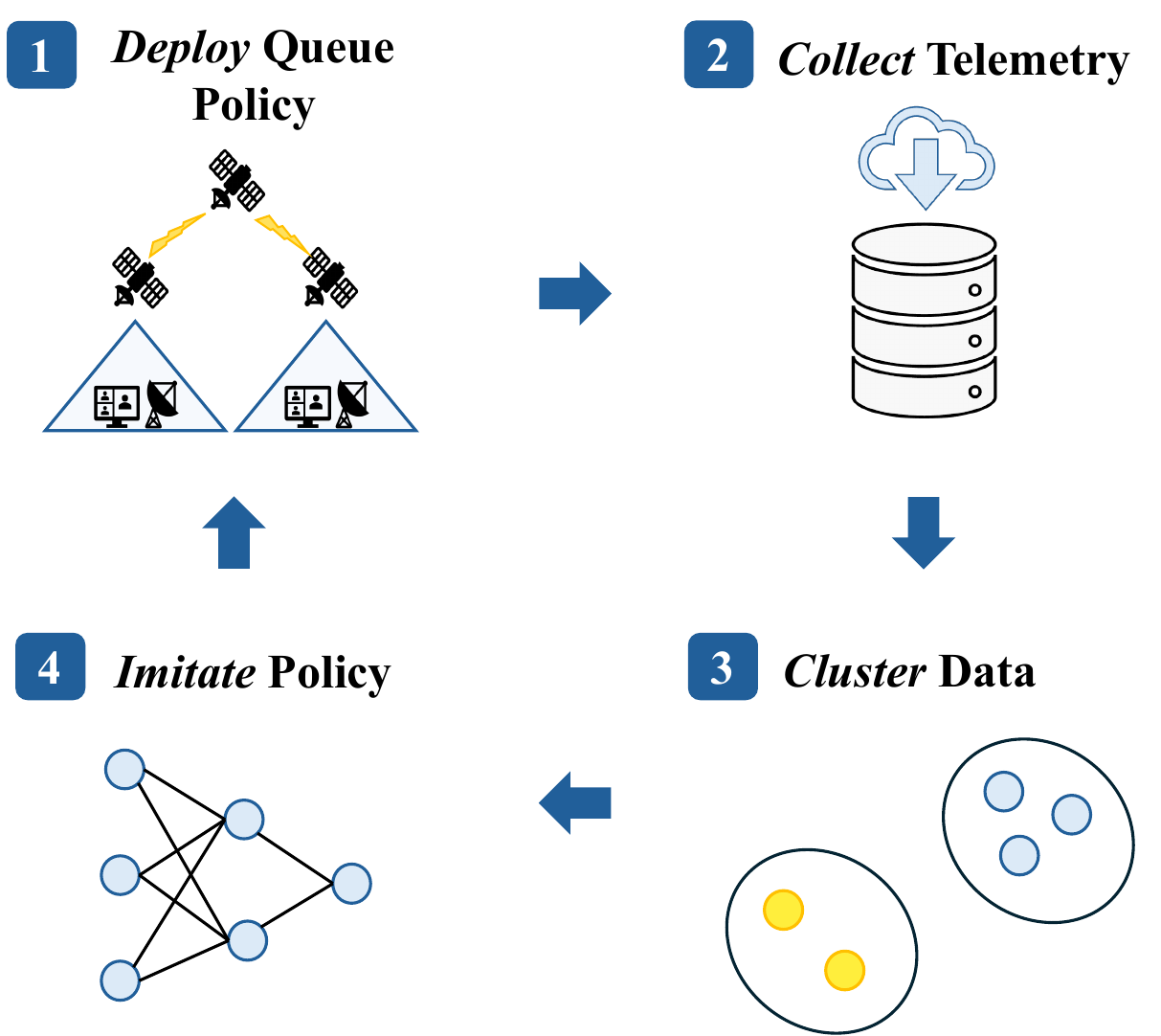}
    \caption{System Workflow: (1) An initial policy is deployed client-side. (2) Telemetry logs are aggregated and (3) clustered to build an expert target policy. (4) A final policy is trained to imitate this expert and is subsequently deployed.}
    \vspace{-5mm}
    \label{fig:architecture}
\end{figure}

Figure~\ref{fig:architecture} describes our system's workflow. First, an initial queue policy is deployed on client devices. Second, we collect telemetry logs generated from these user videoconferencing calls. Client logs are then transferred to a central server and processed to extract pairs of queue policy actions and their corresponding network state variables (e.g., handover metrics). Third, we cluster the resulting state-action pairs based on the number and frequency of observed handovers per decision interval. The target management policy is then constructed by labeling each cluster with the queue limit that produced the highest call quality. This process generates state-action pairs that reflect the experiences of a dynamic queue management policy (\ref{sec:methods_data_collection}). Lastly, our system learns to imitate the target policy from these offline logs (\ref{sec:methods_architecture}) and subsequently deployed on client devices. We describe our implementation in Section~\ref{sec:methods_policy_deployment}.

\subsection{Problem Formulation}
\label{sec:methods_problem_formulation}
Given a fixed set of offline telemetry logs, we seek to learn a
queue management policy that selects the most suitable queue size to utilize for a given decision interval.

{\bf Imitation Learning.} We formulate our learning problem as an offline contextual bandit. Since our decisions are driven by exogenous variables (e.g.,\ satellite handovers) and the agent's actions do not influence future states, the bandit formulation is a more appropriate policy learning formulation as opposed to alternative methods such as RL. Contextual bandit problems are formulated as $(\mathcal{X}, \mathcal{A}, \mathcal{P}(X), r(X,A))$, where $\mathcal{X}$ is the context space (e.g., network states), $\mathcal{A}$ is the action space, $\mathcal{P}(X)$ is the distribution from which contexts are observed at each step, and $r(X,A)$ is the expected reward for taking action $A$ given context $X$. In the offline setting, we are provided with a fixed dataset of experiences, $\mathcal{D} = \{(x_i, a_i, r_i)\}_{i=1}^N$, and cannot interact with or query the environment for new information. This contrasts with online bandit algorithms that can directly query the target environment for exploration and exploitation. Moreover, online IL approaches are infeasible: our expert is not available for interactive queries, but a mapping derived offline from clustered QoE outcomes. Therefore, our goal is to learn a policy $\hat{\pi}: \mathcal{X} \rightarrow \mathcal{A}$ from this static offline dataset.

Rather than directly estimating the reward function $r(X,A)$, our approach leverages IL. We seek to imitate a non-linear queue management heuristic, which acts as our expert. Specifically, given a static dataset of collected demonstrations, $\Xi = \{(s_j, \pi_{\text{exp}}(s_j))\}_{j=1}^{N}$, where $s_j \in \mathcal{X}$ represents the observed network state and $\pi_{\text{exp}}(s_j)$ is the action taken by the expert in state $s_j$, our objective is to learn an imitation policy $\hat{\pi}$. This is achieved by framing the problem as a supervised learning task:
\begin{equation} \label{eq:imitation_objective_bandit}
\hat{\pi}^* = \operatorname*{arg\,min}_{\hat{\pi}} \sum_{s_j \in S} L(\hat{\pi}(s_j), \pi_{\text{exp}}(s_j)) \tag{3}
\end{equation}
where $L$ is a suitable loss function (e.g., cross-entropy for discrete actions) that measures the discrepancy between the imitator policy's action $\hat{\pi}(s_j)$ and the expert heuristic's action $\pi_{\text{exp}}(s_j)$ for a given state $s_j$. $\hat{\pi}^*$ is the learned policy that best mimics the expert based on the provided demonstrations.

\begin{figure}[t!]
    \centering
    \begin{subfigure}[b]{0.48\columnwidth}
        \includegraphics[width=\linewidth, trim={0cm, 0cm, 0cm, 0cm}, clip]{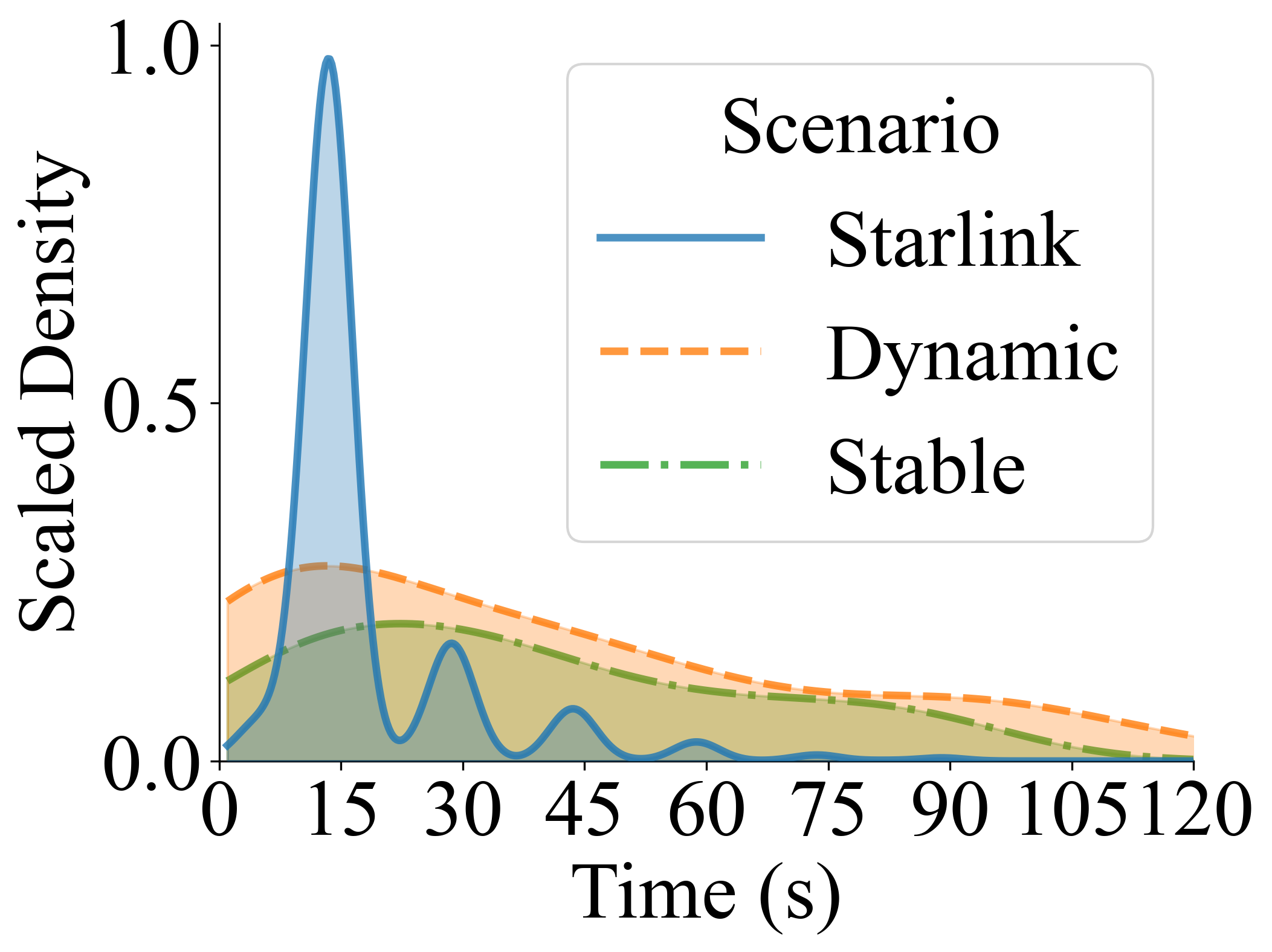}
        \caption{Inter-handover time.}
        \label{fig:handover-distributions}
    \end{subfigure}
    \hfill 
    \begin{subfigure}[b]{0.48\columnwidth}
        \includegraphics[width=\linewidth, trim={0cm, 0cm, 0cm, 0cm}, clip]{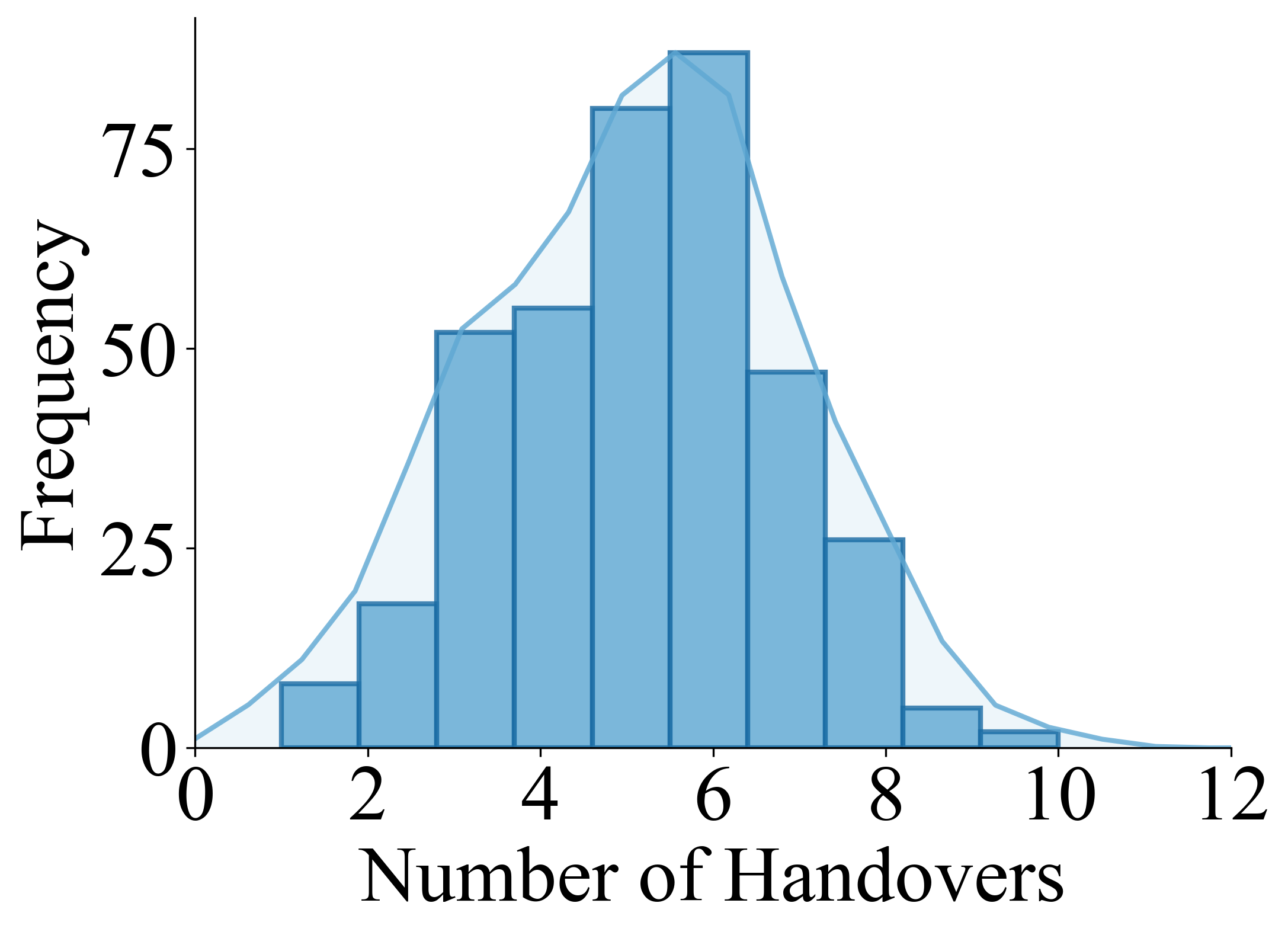}
        \caption{Handovers per window.}
        \label{fig:handover-counts}
    \end{subfigure}
    \caption{Handover characteristics in LEO deployments. (a) Distribution of time between handovers. (b) Distribution of handover counts per 2-minute window in Starlink.}
    \vspace{-5mm}
    \label{fig:handover-analysis}
\end{figure}

{\bf Contextual Bandit.} At the start of decision interval $j$, we define the state $s_j = (\vec{h_j}, \vec{t_j})$ to represent anticipated handover conditions. $\vec{h_j}$ is a binary predictive feature where $\vec{h_{j,t}}=1$ if a handover is likely to occur at time $t+1$, while $\vec{t_j}$ denotes the normalized total expected handovers. These features provide a coarse per-second view of the LEO environment and can be inferred from periodic satellite movement and tools such as starlink-grpc or orbital models (e.g., SGP4 with TLE data)~\cite{ahangarpour2024trajectory}. Capturing the full interval allows our bandit to model time-varying features such as the distribution of handovers (see Figure~\ref{fig:handover-distributions}). We use a $2$-minute horizon to accommodate both sparse and dense constellations, with ablations against $1$-minute and $30$-second horizons validating our design decision. Thus, each state contains $240$ features ($120$ each for $\vec{h_j}$ and $\vec{t_j}$). Given $s_j$, our policy $\pi_\theta$ selects a queue size $a_j$ for the entire interval.

\subsection{Data Collection \& Processing}
\label{sec:methods_data_collection}
To emulate sparse LEO constellations, we utilize StarryNet~\cite{lai2023starrynet}, a high-fidelity container-based ISTN emulator. StarryNet integrates crowd-sourced satellite TLEs and the SGP4 orbital propagation model to capture time-varying dynamics such as satellite positions, propagation delays, visibility, and connectivity. To capture a wide range of realistic network conditions, we emulate sparse LEO constellations ranging from $80$ to $100$ satellites-- with inter-satellite link (ISL) speeds of $1$ Gbps and a packet loss of $1\%$. We additionally vary the location of the user terminals (UT) by sampling coordinates from popular cities across the globe and perturb orbital parameters such as altitude and phasing to emulate satellite maneuvers.

Offline learning is limited by the policies represented in the dataset. To ensure coverage, we collected approximately $65$ hours of WebRTC videoconferencing logs with GCC and a random queue policy in emulated LEO environments, and $12.5$ hours of Starlink logs over one week under varying weather conditions. Live Starlink sessions used a Macbook Pro (via a Starlink Mini) in Austin, Texas connected to a Cloudlab server~\cite{duplyakin2019design} located in Salt Lake City, Utah. Each interval uniformly selected one of 15 queue limits ranging from $100$ to $4000$ ms. We extended WebRTC's telemetry (e.g.,\ packet loss, bitrate, and freeze rate) to capture handover features, queue actions, and QoE video metrics. StarryNet handovers were identified via serving satellite ID changes, while for Starlink we prototyped a handover detector using antenna obstruction maps~\cite{ahangarpour2024trajectory} available over starlink-grpc to capture per-second handover traces. StarryNet employs a delay-based handover policy, intiating a handover once a lower candidate delay is detected, while Starlink's policy is not public. Experience was collected once before training.

To build the target policy, we applied K-means clustering on total number and frequency of handovers to group intervals with similar dynamics (see Figure~\ref{fig:handover-distributions}). Each cluster was labeled with the queue size giving the best QoE, defined as $r_j = 2 * X_j - F_j$ where $X_j$ and $F_j$ correspond to the normalized average video bitrate and video freeze rate observed during decision interval $j$. This simple formulation captures the critical trade-off between throughput and disruptions in videoconferencing. From an initial A/B test of $15$ queue sizes ranging from $100$ to $4000$ ms, we empirically selected $k=5$ clusters, yielding a final action space of five queue limits: $\{500, 600, 900, 2000, 4000\}$ ms.

\subsection{Architecture \& Training} 
\label{sec:methods_architecture}
The policy network was implemented as a two-layer transformer encoder and trained directly from offline logs. Each time step was tokenized into a $64$ dimensional embedding using a multi-layer perceptron (MLP). We set the batch size to $128$ and minimized the cross entropy loss between our policy's actions and expert actions over the course of $100$ training epochs. We employed Adam optimization with a cosine annealing learning rate schedule with the following parameters: initial learning rate of $0.001$, a weight decay of $0.01$, and a dropout rate of $0.1$ for regularization. Hyperparameter selection was validated using five-fold cross-validation.

\subsection{Policy Deployment}
\label{sec:methods_policy_deployment}
We implemented our queue selection policy in Pytorch and integrated it directly on top of WebRTC. Specifically, our queue selection policy runs on end-devices and directly modifies the maximum queue limit of WebRTC's send-side packet pacer through the pacer's API (e.g.,\ {\it setQueueLimit(TimeDelta limit)}). Additionally, given that our policy executes in near real-time (on the order of seconds), we are not limited to real-time constraints (e.g.,\ sub $50$ ms). In emulation, we leverage predictable orbital motion of satellite constellations, SGP4, and TLE data to estimate high-fidelity LEO features for the upcoming decision interval. In contrast, for our Starlink deployment, privileged details such as the serving satellite ID are not publicly accessible. To estimate handovers, we assume patterns are consistent across consecutive decision intervals-- an assumption based on the predictable orbits of satellites~\cite{tanveer2023making, ahangarpour2024trajectory}. Consequently, the handover events logged by our handover detector in the previous interval serve as the estimate for the upcoming one. For the initial interval, where prior data is unavailable, we default to an estimate of $8$ equally-spaced handovers. We leave more advanced handover prediction as an area of future work. 
\section{Evaluation}
\label{sec:eval}

The key takeaways from our evaluations are as follows:
\begin{enumerate}
\item Our learned policy did not produce a statistically significant difference in key performance indicators when compared to WebRTC's default pacing queue policy in emulated terrestrial settings. 
\item Our \textit{handover-aware} queue policy delivers up to a $3$x increase in video bitrate, while reducing the video freeze rate by up to $62\%$ in comparison to WebRTC's default queue policy in emulated sparse constellations.
\item Our policy dynamically adapts the maximum send-side pacing queue to absorb handover disruptions, delivering up to a $41\%$ reduction in freeze rate and $40\%$ decrease in mean packet loss on real Starlink constellations.
\end{enumerate}

\subsection{Testbed Setup}
Our queue policy was evaluated with peer-to-peer videoconferencing calls conducted within a controlled testbed and real satellite environments. Our setup leveraged Cloudlab~\cite{duplyakin2019design} for the underlying compute infrastructure, AlphaRTC~\cite{eo2022opennetlab}-- an open-source WebRTC fork-- as the videoconferencing client, and the StarryNet network emulation tool~\cite{lai2023starrynet}. StarryNet was configured using publicly available TLE data to emulate dynamic satellite network conditions. For live deployments, we utilized Starlink as our network provider. We primarily evaluated our approach (Ours) against WebRTC's default queue limit policy (Def), where GCC served as the common rate control algorithm for both setups. We report the following key performance metrics: video bitrate (Mbps), freeze rate (freezes/minute), video frame rate (frames/second), P95 end-to-end delay (ms), and the mean packet loss (\%). We perform A/B testing across hours of live calls and report our results below.

\subsection{Terrestrial Networks}
\begin{figure*}[ht!]
\centering
\begin{subfigure}[b]{0.19\textwidth}
\centering
\includegraphics[width=\textwidth, trim={0 0cm 0 0cm}, clip]{"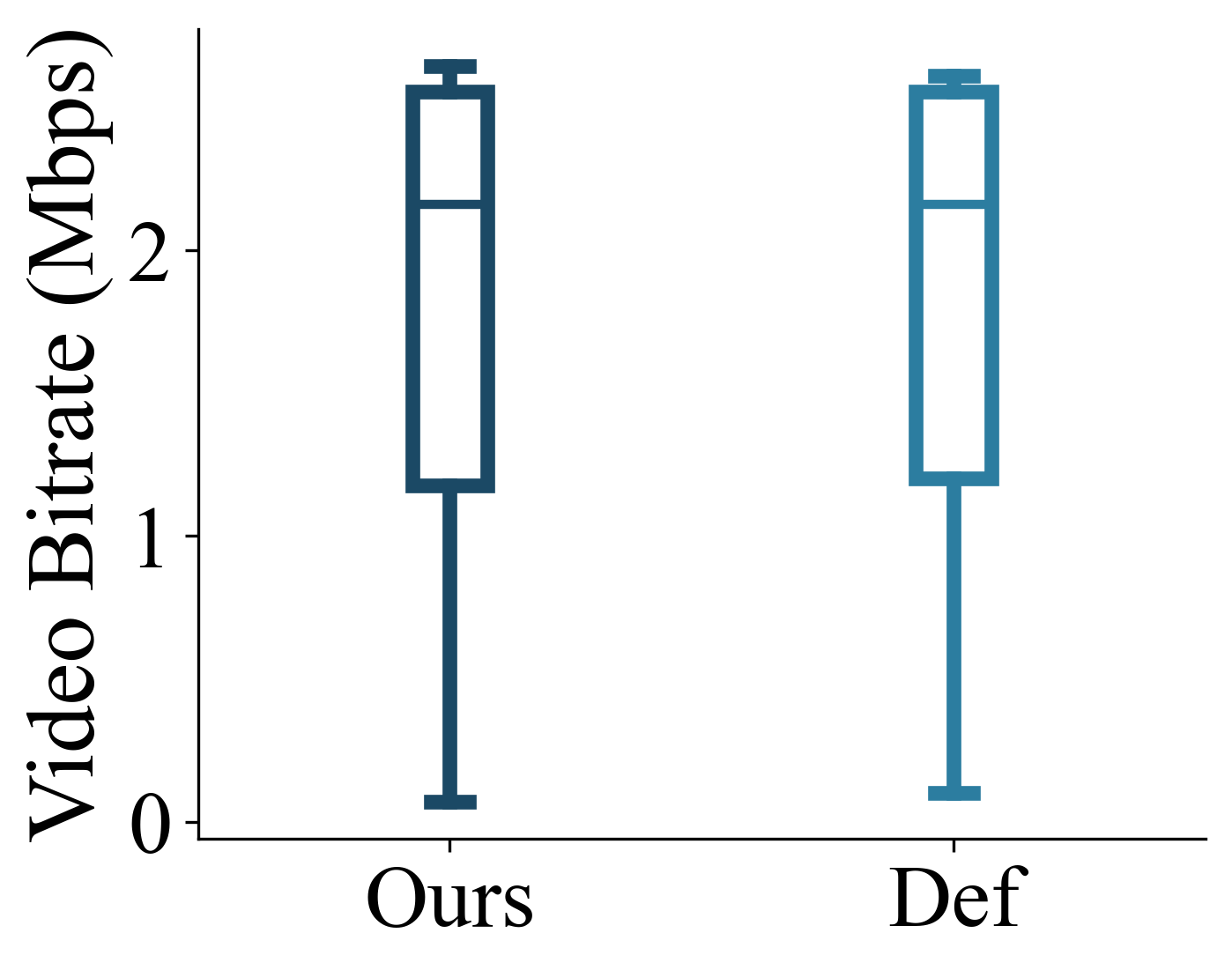"}
\caption{Video Bitrate.}
\label{fig:terrestrial-bitrate}
\end{subfigure}
\hfill
\begin{subfigure}[b]{0.19\textwidth}
\centering
\includegraphics[width=\textwidth, trim={0 0cm 0 0cm}, clip]{"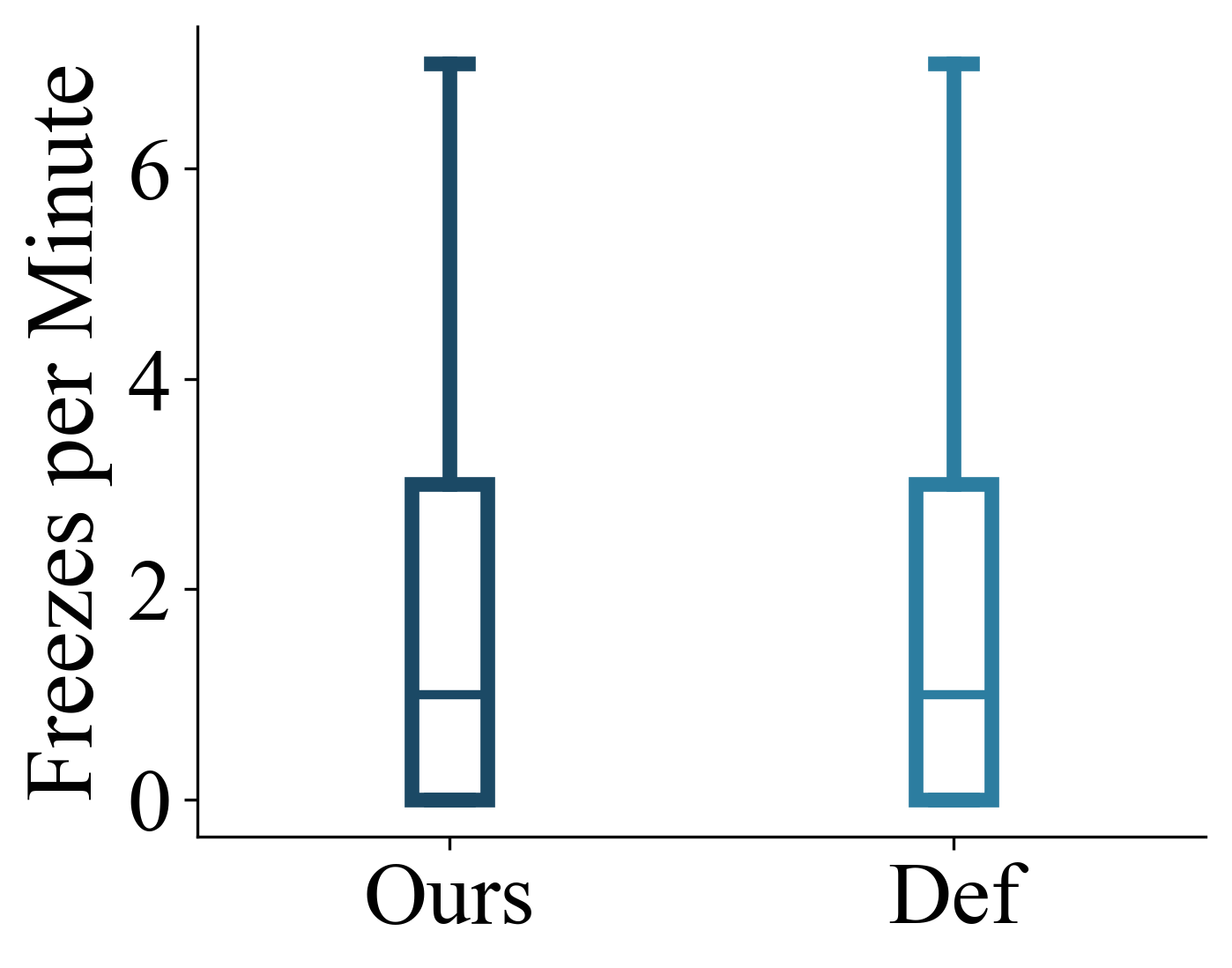"}
\caption{Video Freeze Rate.}
\label{fig:terrestrial-freeze}
\end{subfigure}
\hfill
\begin{subfigure}[b]{0.19\textwidth}
\centering
\includegraphics[width=\textwidth, trim={0 0cm 0 0cm}, clip]{"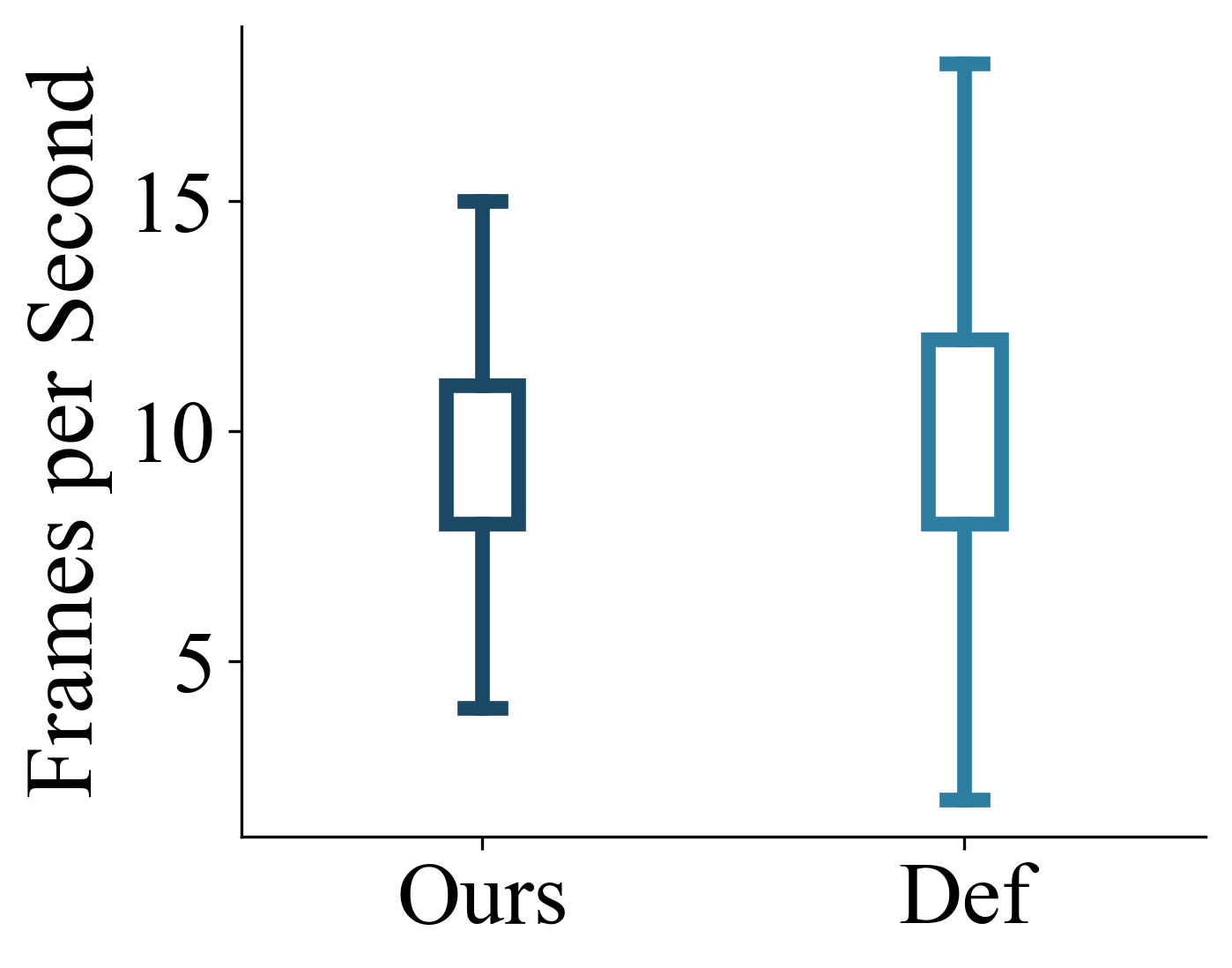"}
\caption{Video Frame Rate.}
\label{fig:terrestrial-fps}
\end{subfigure}
\hfill
\begin{subfigure}[b]{0.19\textwidth}
\centering
\includegraphics[width=\textwidth, trim={0 0cm 0 0cm}, clip]{"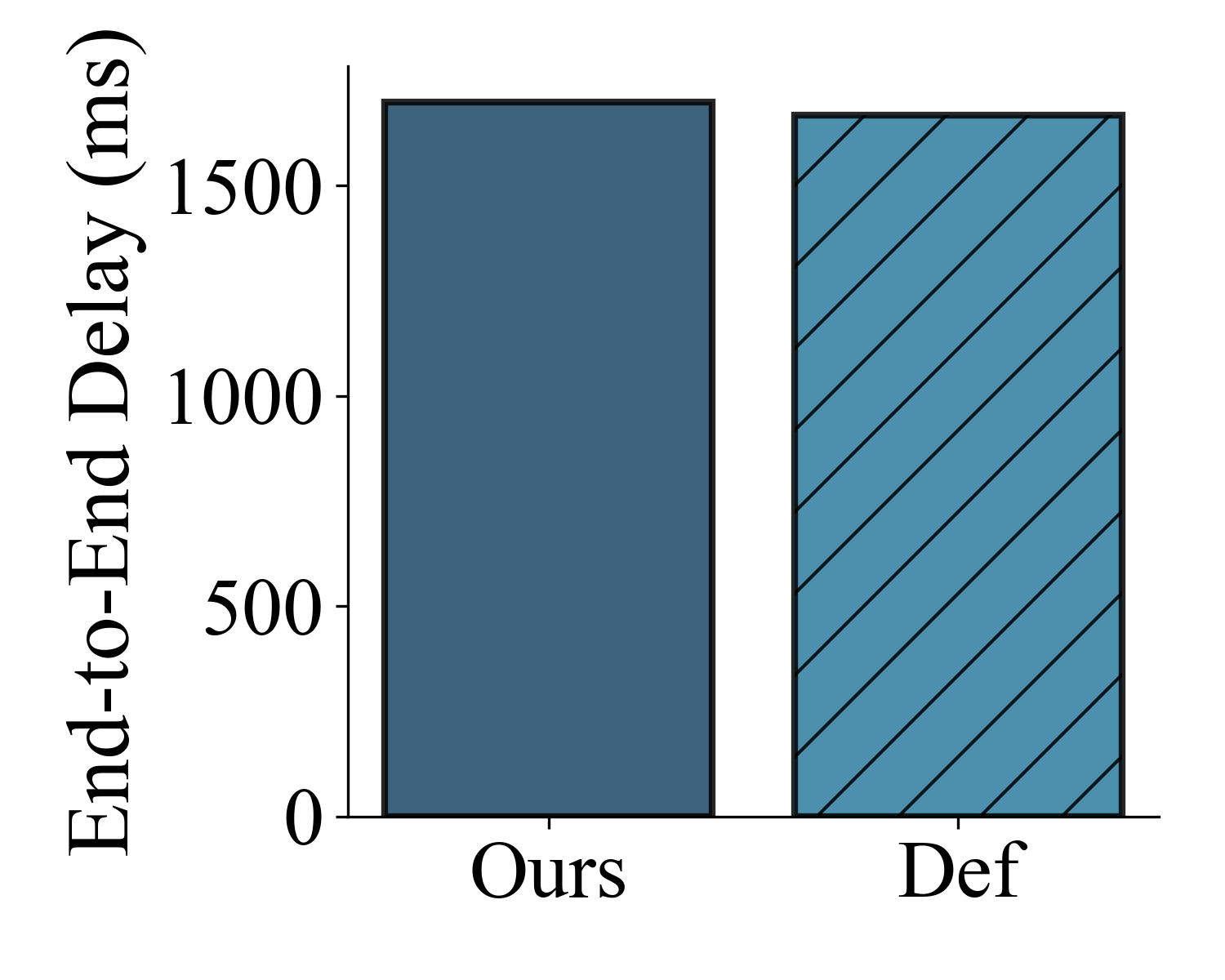"}
\caption{P95 E2E Delay.}
\label{fig:terrestrial-delay}
\end{subfigure}
\hfill
\begin{subfigure}[b]{0.19\textwidth}
\centering
\includegraphics[width=\textwidth, trim={0 0cm 0 0cm}, clip]{"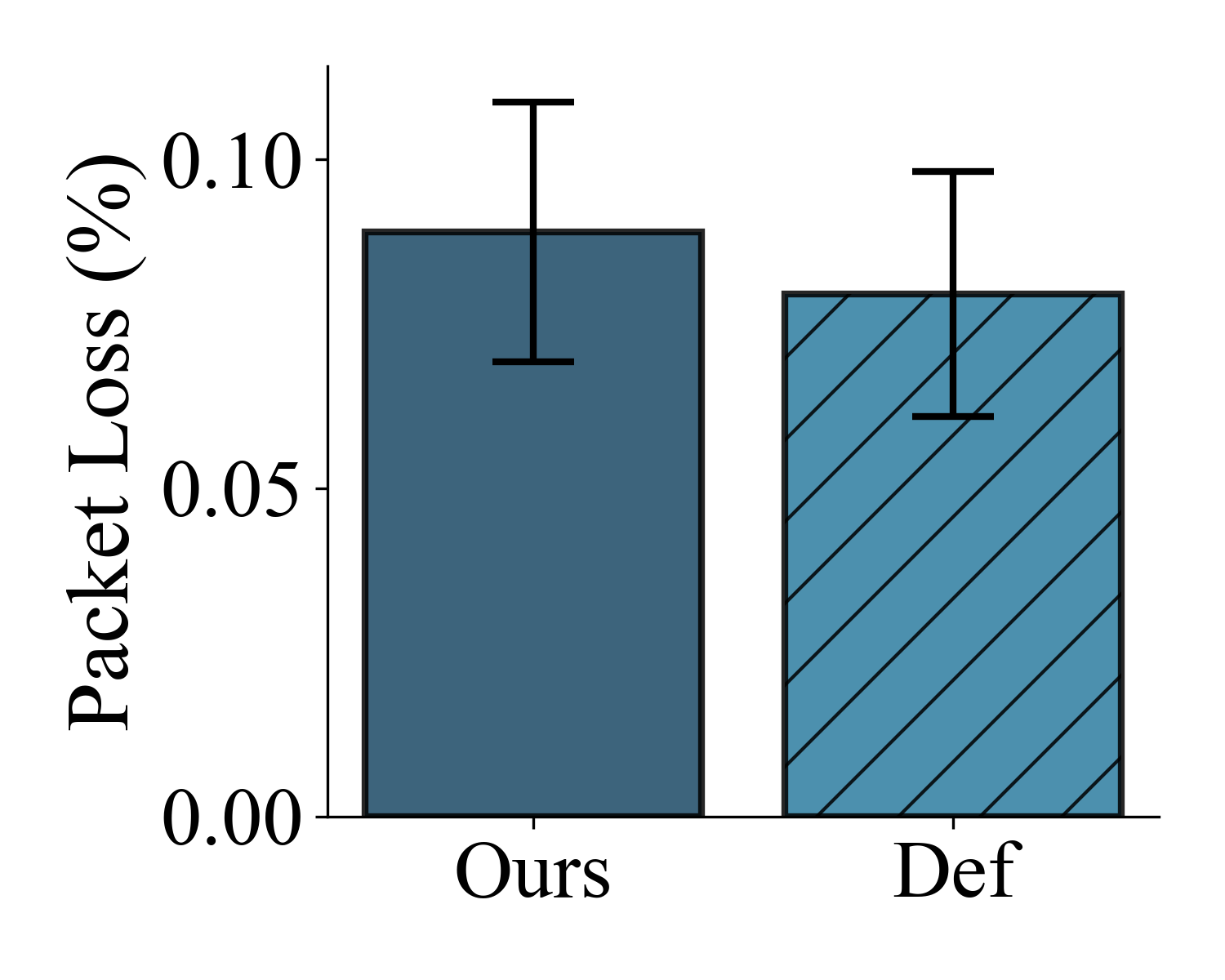"}
\caption{Mean Packet Loss.}
\label{fig:terrestrial-loss}
\end{subfigure}
\caption{On emulated terrestrial links, we observe no impact on key metrics, which suggests that dynamic pacing (Ours) offers little benefit over the default policy (Def) in these low-latency environments.}
\vspace{-5mm}
\label{fig:terrestrial}
\end{figure*}

In this experiment, we seek to demonstrate that our queue policy does not negatively impact terrestrial sessions. We baseline our learned queue policy against GCC's default queue policy. We utilized publicly available 5G, 4G, and broadband traces~\cite{he2024designing} and Mahimahi~\cite{netravali2015mahimahi} to emulate terrestrial settings. As handover information is unavailable in this environment, our learning-based method accepts a zero vector as input. Our evaluation consisted of over $15$ hours worth of videoconferencing calls, we report our key metrics in Figure~\ref{fig:terrestrial}.

When supplied with a zero-vector input (representing no handovers), we observe that our model defaults to a $500$ ms queue size. This is the intended behavior; the absence of handovers indicates a stable network where the policy's objective is to maximize the transmission rate. Despite this, our learning-based approach showed no statistically significant difference in performance compared to WebRTC's default policy in these terrestrial settings (see Figure~\ref{fig:terrestrial}). This finding supports our hypothesis that dynamic pacing has limited utility in low-latency terrestrial environments and is primarily beneficial for addressing the unique delay dynamics of LEO satellite networks. 

\subsection{Sparse Satellite Constellations}
\begin{figure*}[ht!]
\centering
\begin{subfigure}[b]{0.19\textwidth}
\centering
\includegraphics[width=\textwidth, trim={0 0cm 0 0cm}, clip]{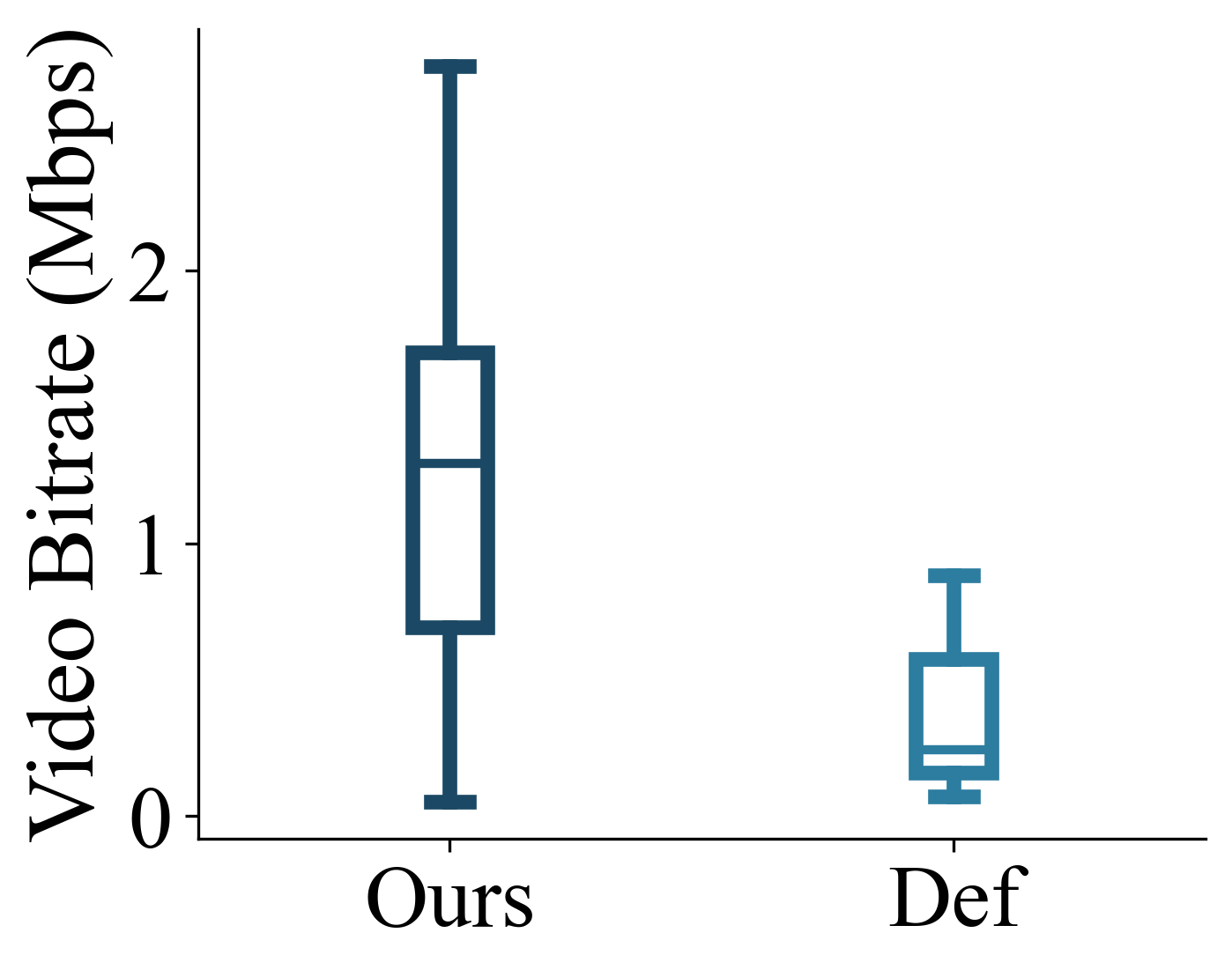}
\caption{Video Bitrate.}
\label{fig:sparse-bitrate}
\end{subfigure}
\hfill
\begin{subfigure}[b]{0.19\textwidth}
\centering
\includegraphics[width=\textwidth, trim={0 0cm 0 0cm}, clip]{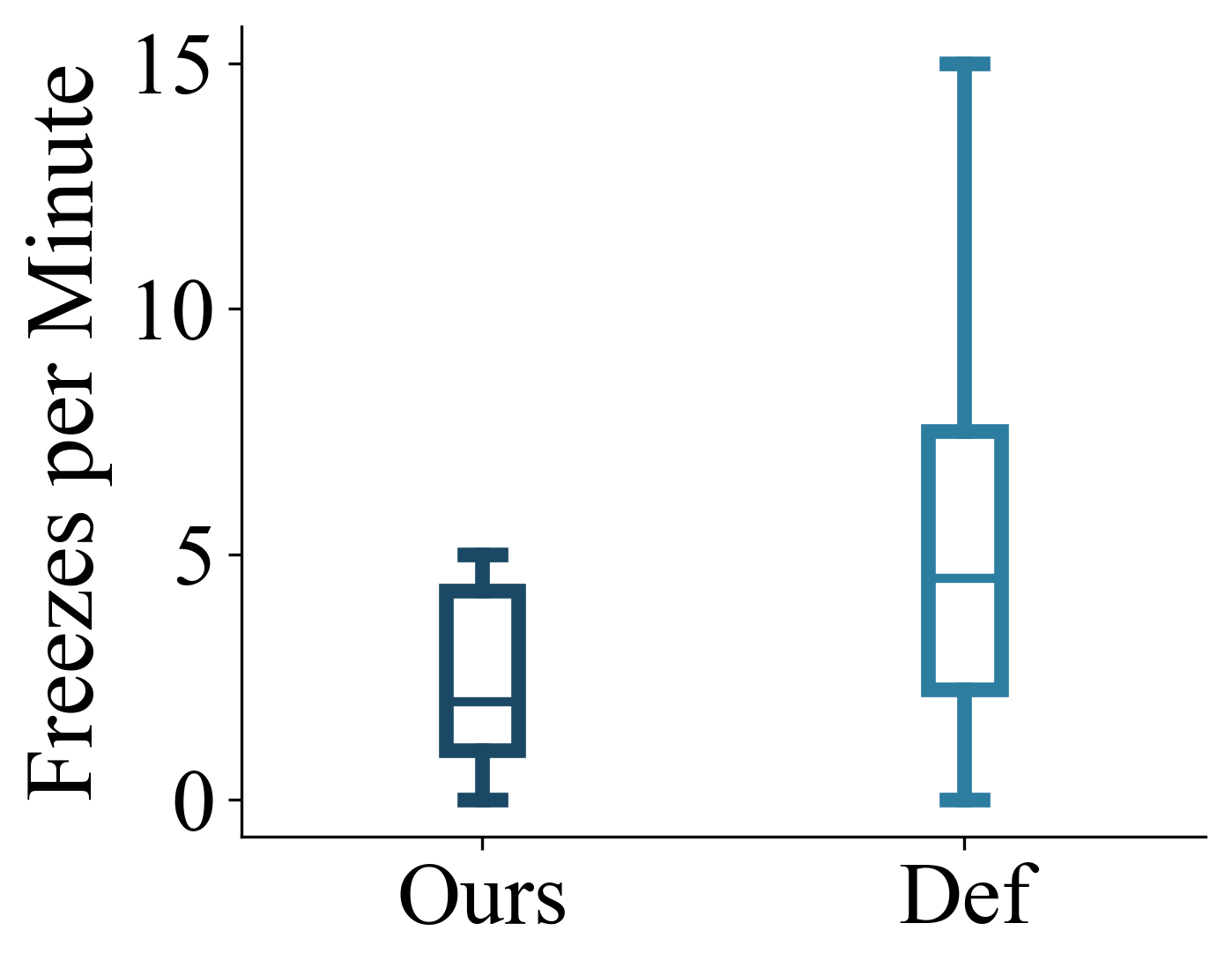}
\caption{Video Freeze Rate.}
\label{fig:sparse-freeze}
\end{subfigure}
\hfill
\begin{subfigure}[b]{0.19\textwidth}
\centering
\includegraphics[width=\textwidth, trim={0 0cm 0 0cm}, clip]{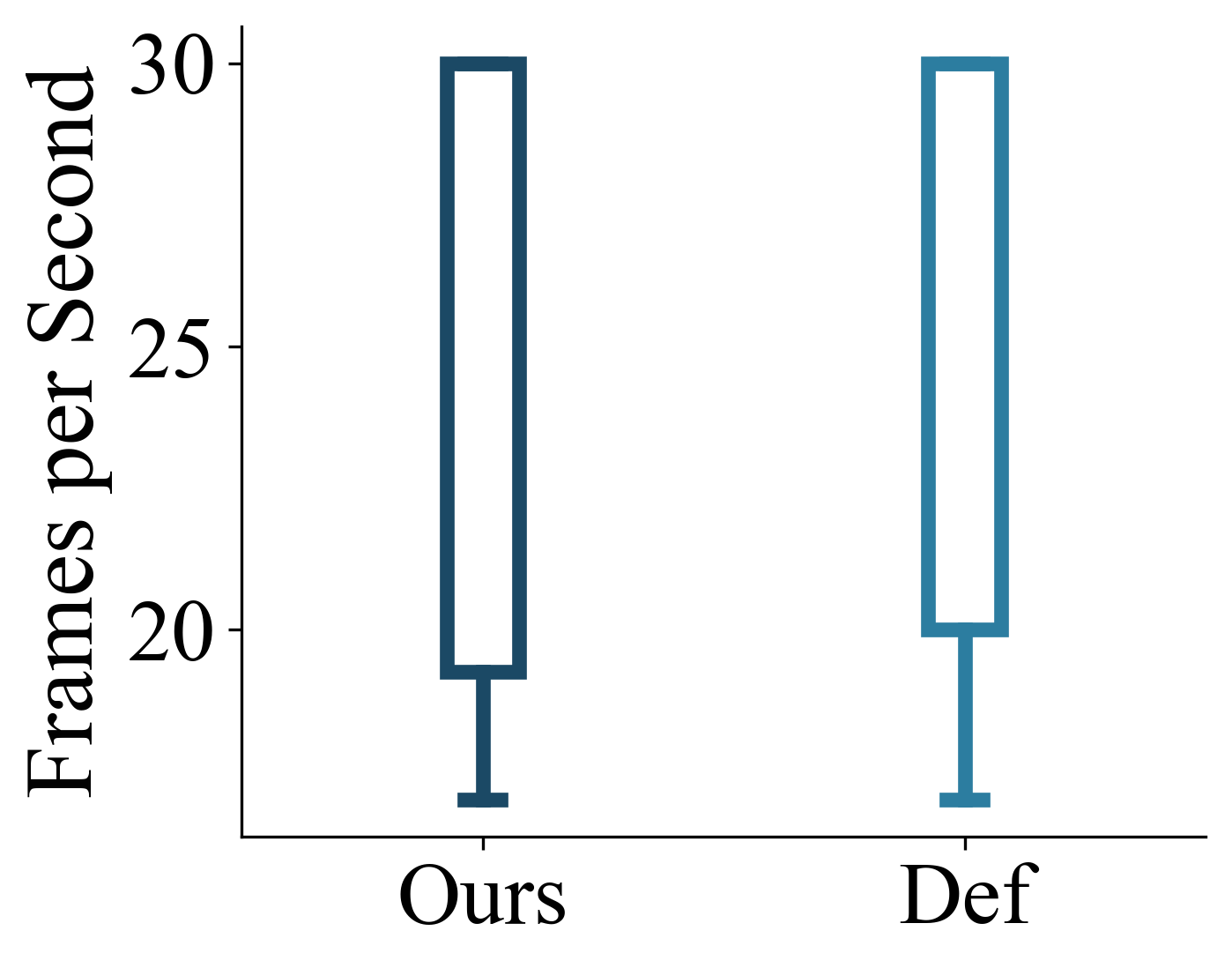}
\caption{Video Frame Rate.}
\label{fig:sparse-fps}
\end{subfigure}
\hfill
\begin{subfigure}[b]{0.19\textwidth}
\centering
\includegraphics[width=\textwidth, trim={0 0cm 0 0cm}, clip]{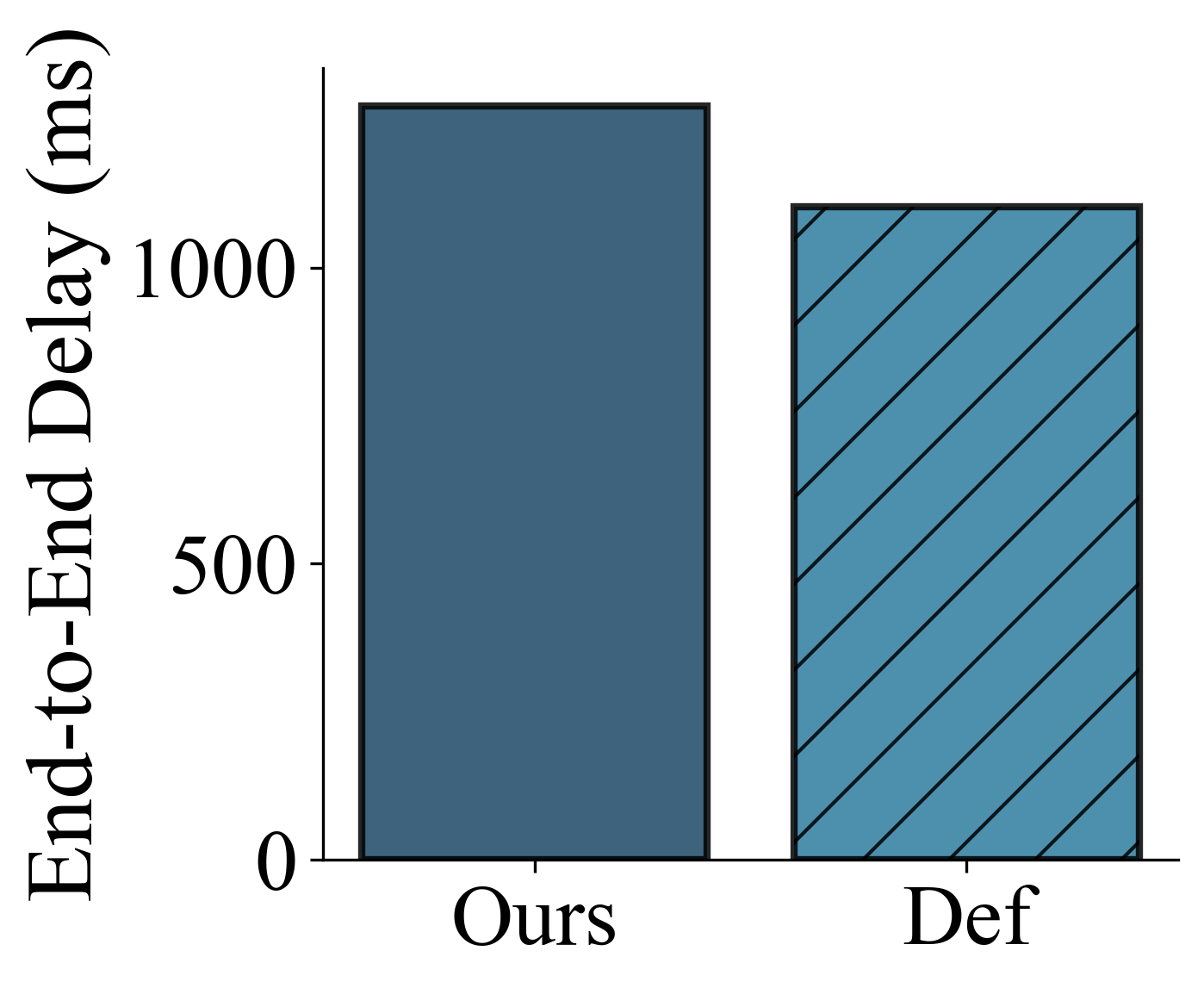}
\caption{P95 E2E Delay.}
\label{fig:sparse-delay}
\end{subfigure}
\hfill
\begin{subfigure}[b]{0.19\textwidth}
\centering
\includegraphics[width=\textwidth, trim={0 0cm 0 0cm}, clip]{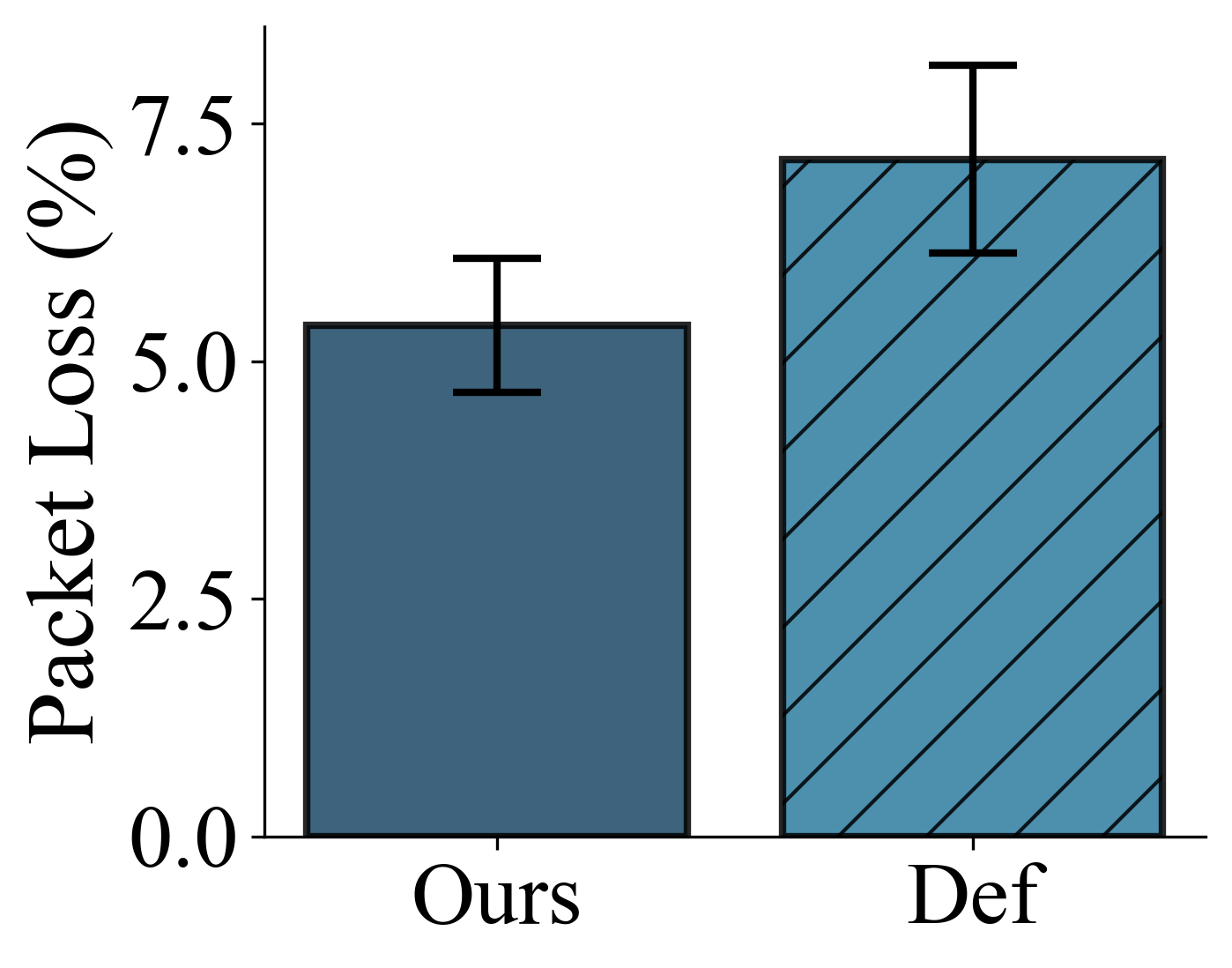}
\caption{Mean Packet Loss.}
\label{fig:sparse-loss}
\end{subfigure}
\caption{In emulated sparse LEO environments, our \textit{handover-aware} pacing policy (Ours) outperforms WebRTC's default (Def), significantly reducing video freezes while improving video bitrate.}
\label{fig:sparse}
\end{figure*}

We baseline our queue policy against GCC's default queue policy. We utilized publicly available TLE data from SpaceTrack to emulate sparse LEO constellations containing $80$ satellites. In this sparse configuration, handovers occurred infrequently, at a rate of approximately $0$-$1$ per minute. For each evaluation call, we randomly sampled pairs of locations from major cities around the globe to represent geographically separated peers. Our evaluation consisted of over $15$ hours worth of videoconferencing calls per queue method. Our key metrics are summarized in Figure~\ref{fig:sparse}.

The results demonstrate that our method generalizes from offline training to an online emulated environment. Our approach leads to a marginal increase in the end-to-end delay; however, we observe no impact on the video frame rate. In contrast, our handover-aware queue policy delivers up to a $3$x increase in video bitrate, while reducing the video freeze rate by up to $62\%$ and the mean packet loss by up to $31\%$ compared to WebRTC's default queue policy (see Figure~\ref{fig:sparse}). We observe that the default queue policy, which allows packets to buffer for up to $2$ seconds, performs poorly in stable LEO conditions. This excessive buffering causes bloated sender-side queues; draining them introduces delay spikes and burst traffic that leads GCC to conservatively lower its target bitrate. As a result, this self-inflicted delay prevents GCC from capitalizing on available network capacity during the low handover periods. In contrast, our proposed policy limits the maximum queue size to $500$-$600$ ms. This aggressive queue policy prevents buildup while ensuring smooth packet flow, and provides GCC with a more accurate view of the network's delay characteristics. As a result, GCC can increase its bandwidth estimates, enabling the video encoder to opportunistically encode frames at a higher quality and bitrate. 

\subsection{Generalizing to Live Deployments}
\begin{figure*}[ht!]
\centering
\begin{subfigure}[b]{0.19\textwidth}
\centering
\includegraphics[width=\textwidth, trim={0 0cm 0 0cm}, clip]{"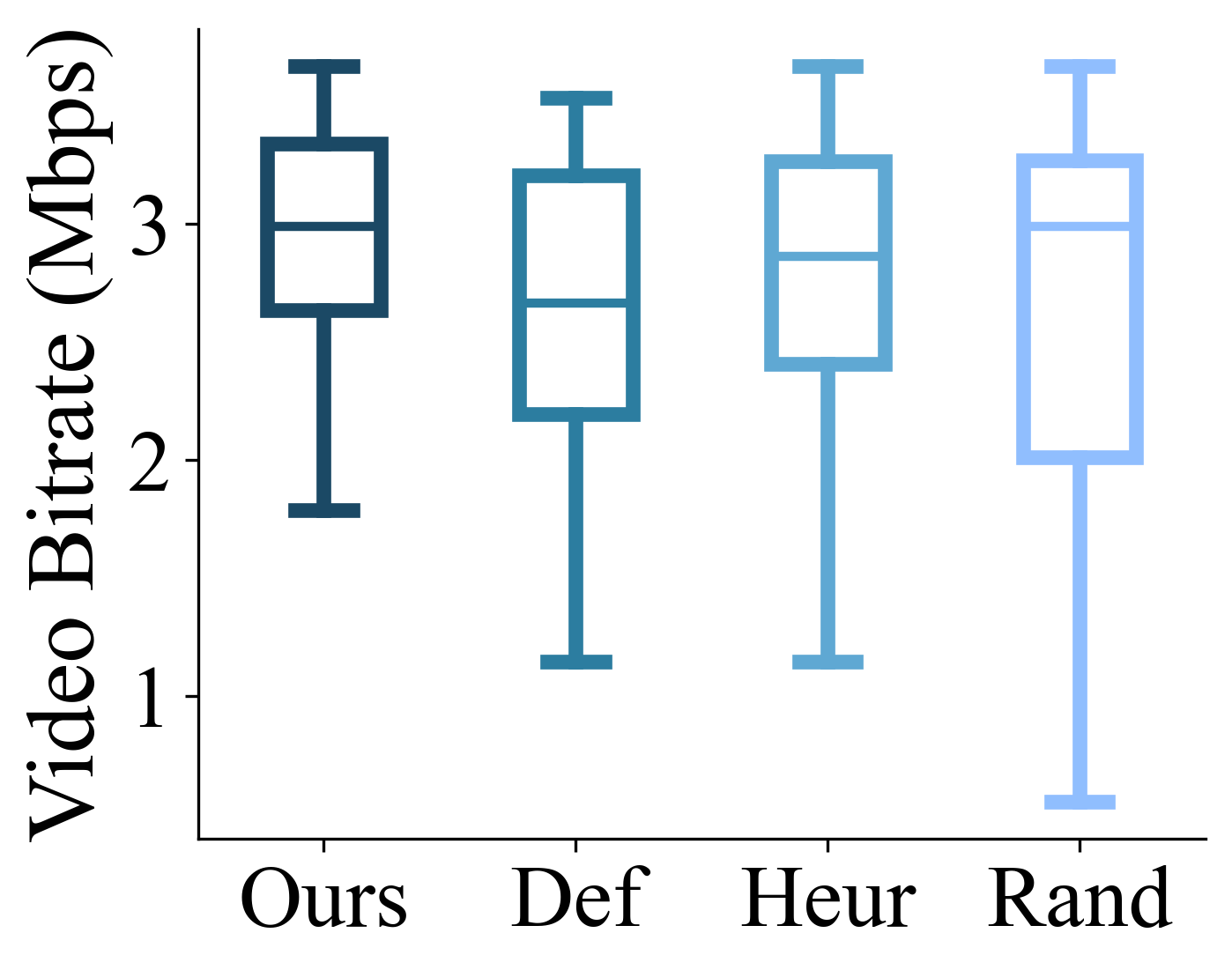"}
\caption{Video Bitrate.}
\label{fig:starlink-bitrate}
\end{subfigure}
\hfill
\begin{subfigure}[b]{0.19\textwidth}
\centering
\includegraphics[width=\textwidth, trim={0 0cm 0 0cm}, clip]{"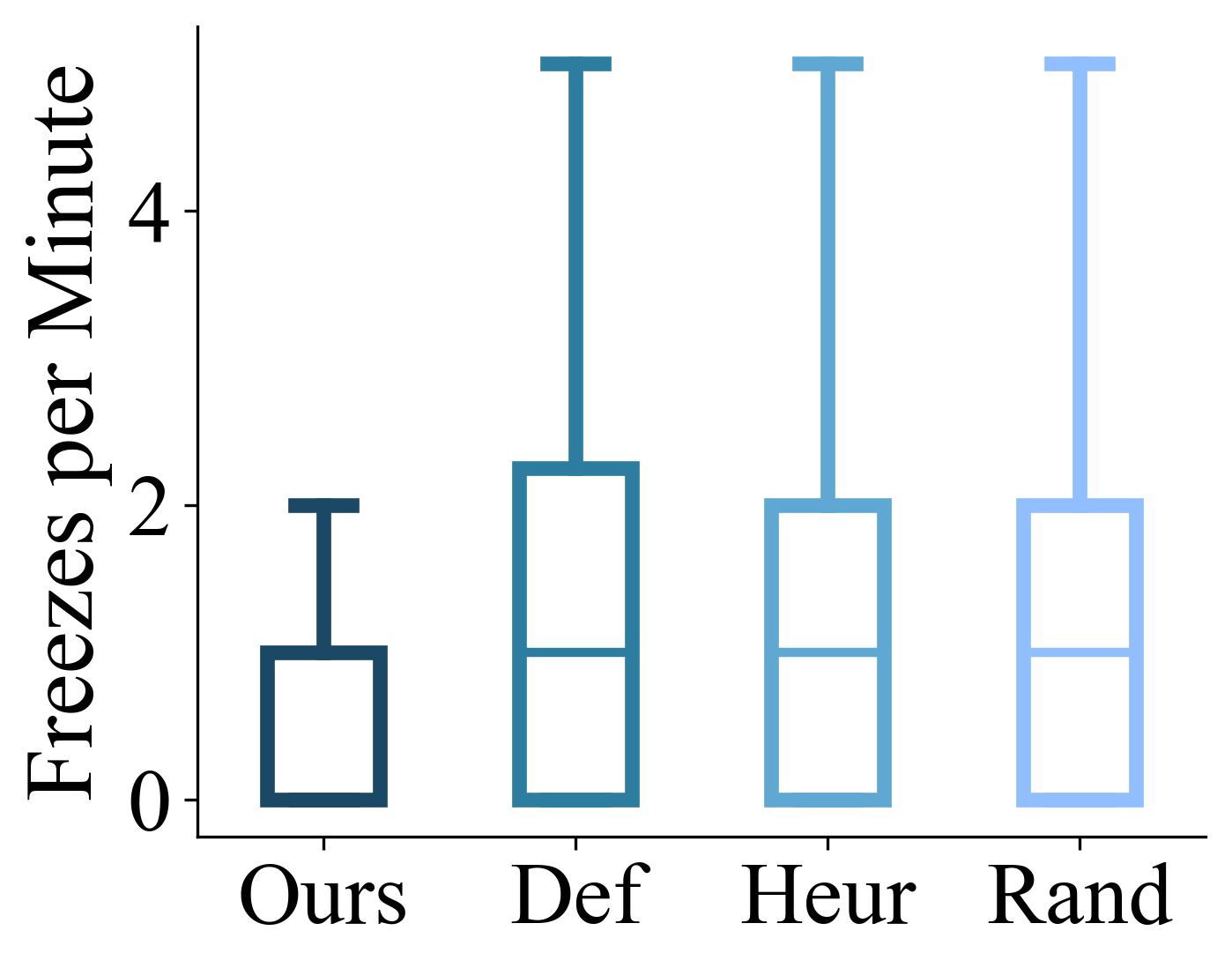"}
\caption{Video Freeze Rate.}
\label{fig:starlink-freeze}
\end{subfigure}
\hfill
\begin{subfigure}[b]{0.19\textwidth}
\centering
\includegraphics[width=\textwidth, trim={0 0cm 0 0cm}, clip]{"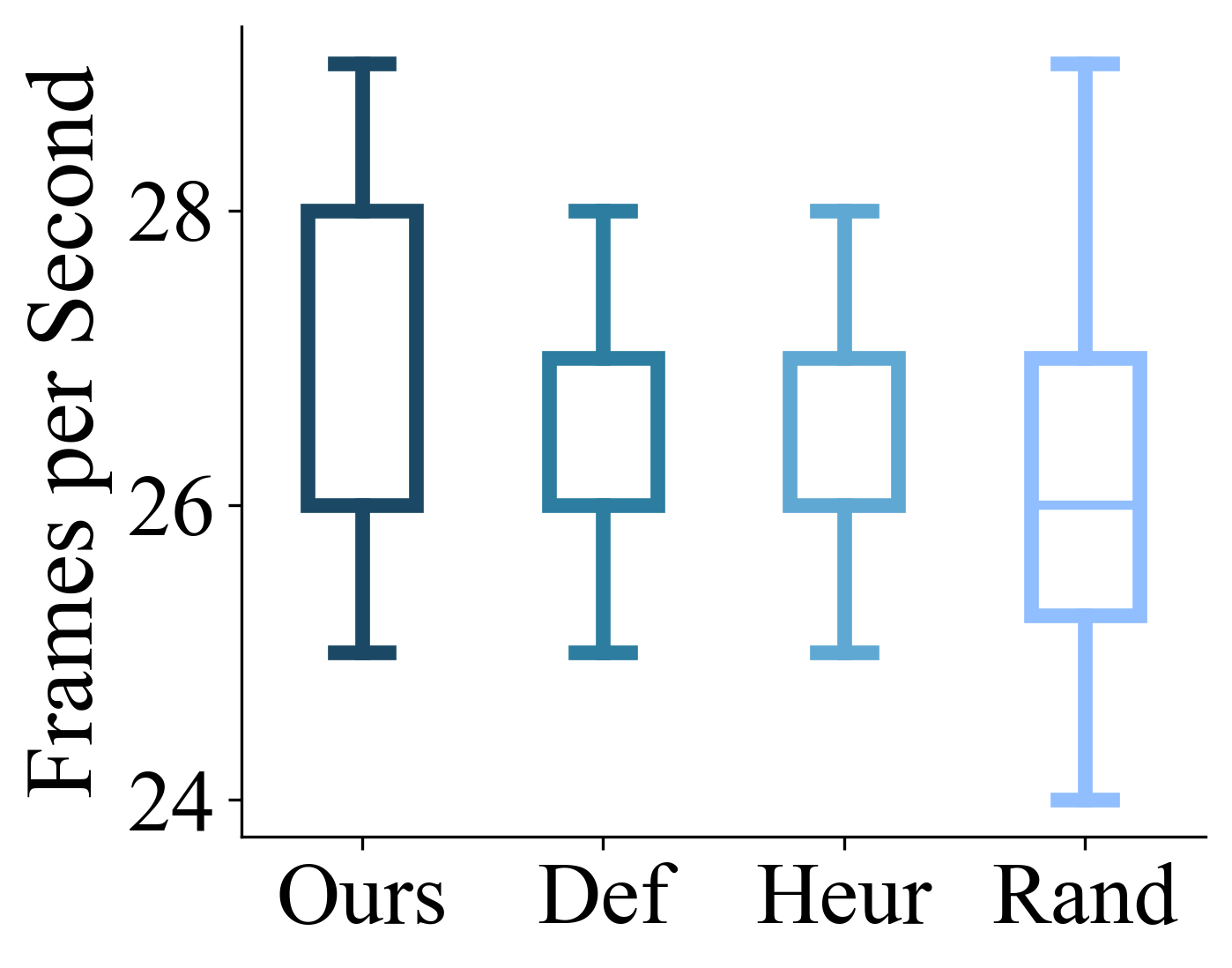"}
\caption{Video Frame Rate.}
\label{fig:starlink-fps}
\end{subfigure}
\hfill
\begin{subfigure}[b]{0.19\textwidth}
\centering
\includegraphics[width=\textwidth, trim={0 0cm 0 0cm}, clip]{"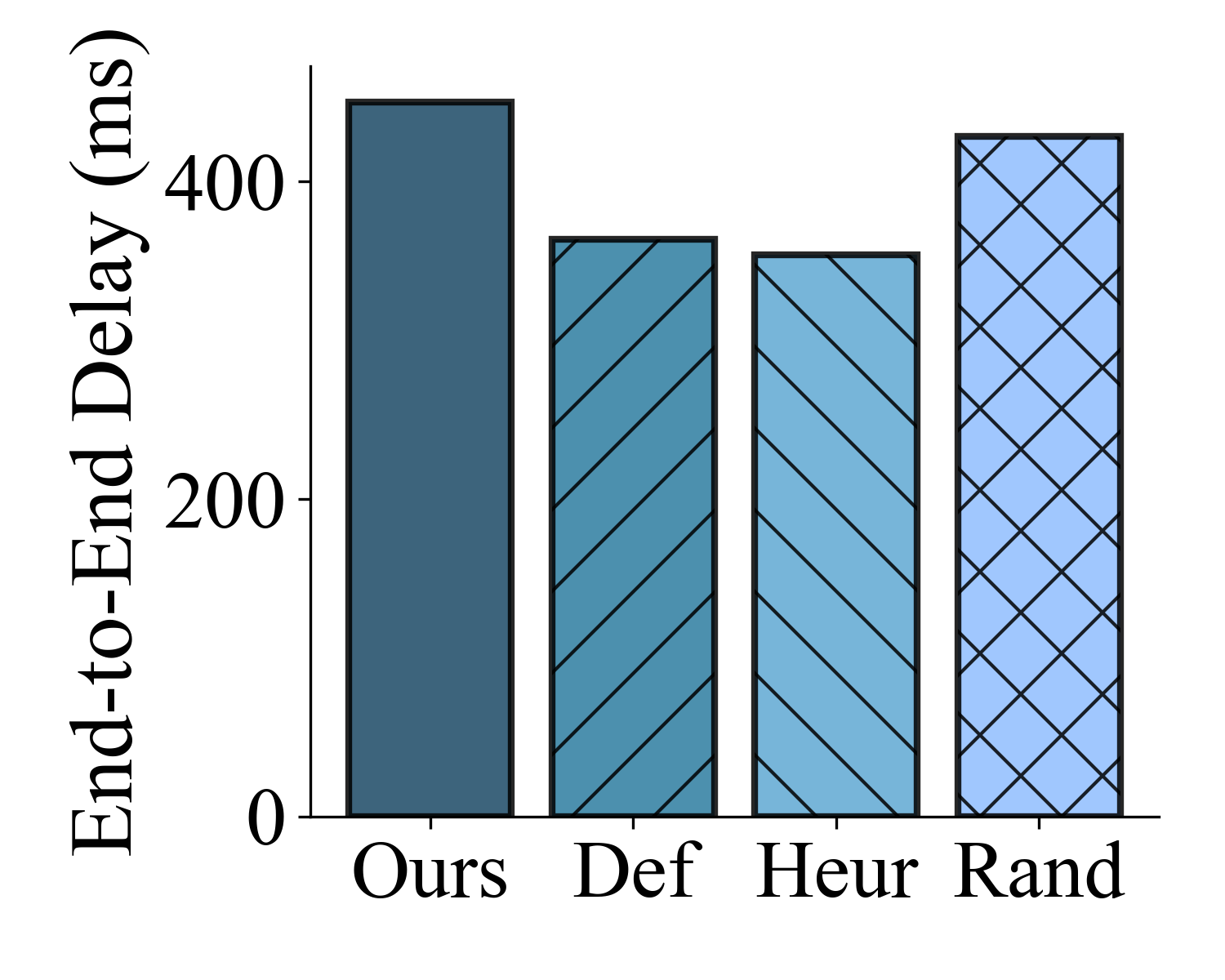"}
\caption{P95 E2E Delay.}
\label{fig:starlink-delay}
\end{subfigure}
\hfill
\begin{subfigure}[b]{0.19\textwidth}
\centering
\includegraphics[width=\textwidth, trim={0 0cm 0 0cm}, clip]{"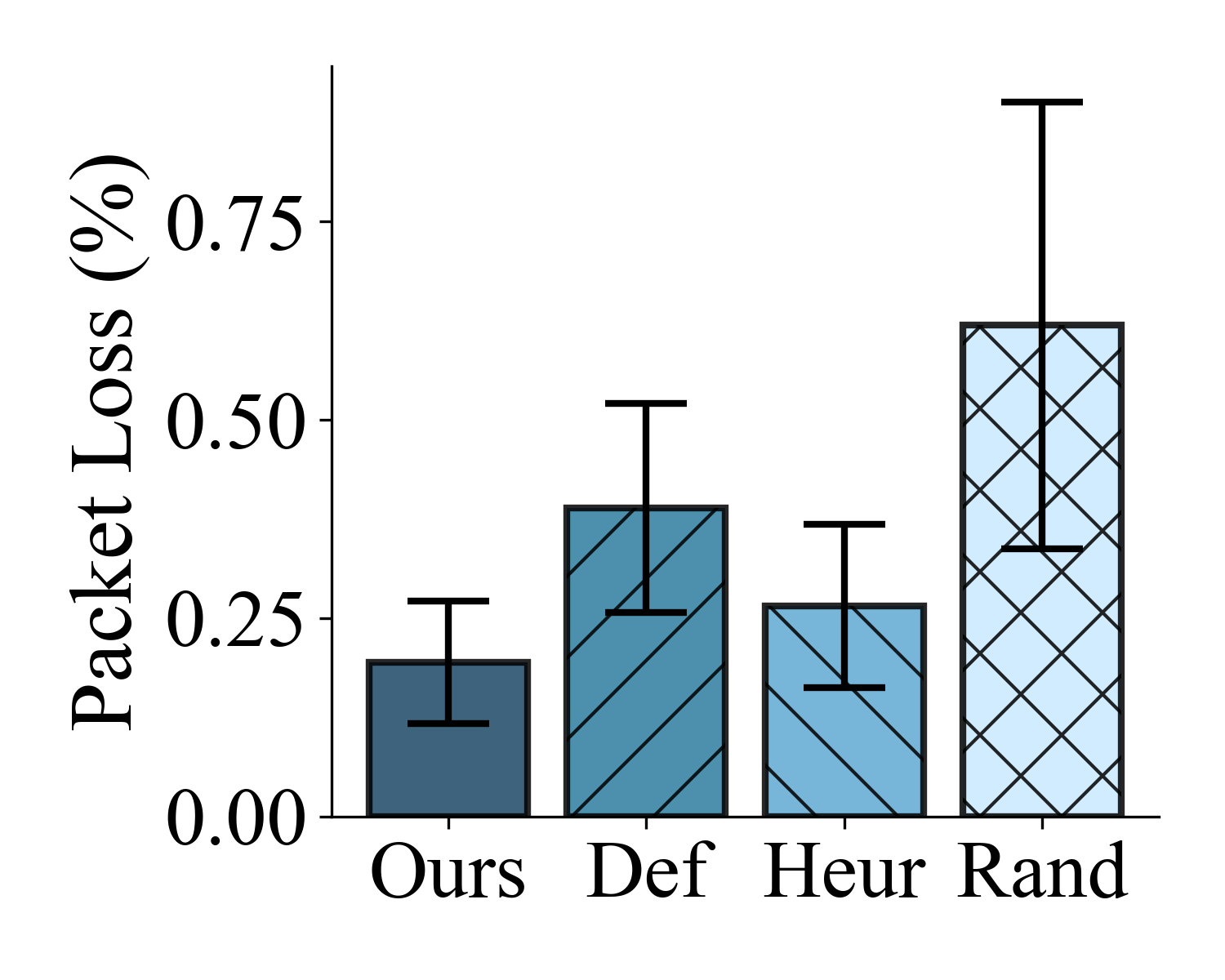"}
\caption{Mean Packet Loss.}
\label{fig:starlink-loss}
\end{subfigure}
\caption{On real Starlink links, our learning-based pacing policy (Ours) outperforms baselines in freezes and packet loss.}
\vspace{-5mm}
\label{fig:starlink}
\end{figure*}

We baseline our queue policy against three baselines on a live Starlink network: (1) the default WebRTC configuration (Def); (2) a policy that randomly selects a queue size at each decision interval (Rand); (3) a rule-based heuristic (Heur). 
This heuristic was derived from offline data analysis and assigns queue limits based on the number of handovers observed in a decision interval. Specifically, it maps $0$, $1$-$2$, $3$-$5$, $6$-$7$, and $8$+ handovers to $500$, $600$, $900$, $2000$, and $4000$ ms respectively. In this experiment, we conducted videoconferencing sessions between a Macbook Pro (tethered to a Starlink Mini) in Texas and a server located in Cloudlab's Emulab cluster in Utah. After following the data collection process detailed in Section~\ref{sec:methods}, we generated a new pacing policy and evaluated our method on a live Starlink network. We conducted $4$ hours worth of videoconferencing sessions, selecting each method for evaluation in a round-robin fashion. Our key metrics are summarized in Figure~\ref{fig:starlink}.

In real-world Starlink deployments, our method continues to outperform the baselines. While performance gains in sparse environments came from improved network resource utilization, the performance gains on Starlink stem from strategically enlarging the sender-side pacing queue to absorb handover-related instability. During intervals with rapid handovers, our method expands the pacing queue to $4000$ ms. This larger buffer allows the sender to absorb disruptions by temporarily holding packets instead of transmitting them on to an unstable link. By minimizing the impact of handover-related disruptions, GCC can better estimate available network resources, leading to a smoother experience.

Compared to the default WebRTC configuration, our method achieves a $41\%$ reduction in the mean freeze rate and a $40\%$ reduction in mean packet loss (see Figure~\ref{fig:starlink}). We hypothesize that avoiding transmissions during satellite handovers prevents packet loss often caused by the route changes and satellite buffer instability inherent to the handover process. Additionally, we observe a marginal increase in the P95 end-to-end delay, which is due to packets strategically waiting in the pacing queue during periods of rapid handovers. Lastly, despite the increase in P95 end-to-end delay, we observe no impact on the video frame rate.

Finally, while our rule-based heuristic improves upon the default WebRTC configuration, it performs significantly worse than our learning-based method. The key weakness of the heuristic is that it neglects the temporal structure of handovers-- it considers only the count of handovers, not their timing within a decision window. As a result, our learned method achieves a $28\%$ reduction in mean freeze rate and a $27\%$ reduction in mean packet loss compared to the heuristic (see Figure~\ref{fig:starlink}).

\section{Limitations}
\label{sec:future}
A key limitation of our work is the reliance on a simplified handover estimator. For live Starlink tests, our current approach predicts handovers for the upcoming decision interval by assuming they will mirror the patterns of the previous one, leading to imprecise estimates. Additionally, it only identifies that a handover is likely to occur, neglecting finer details. For instance, our handover predictor does not distinguish between soft and hard handovers, detect disruptive ping-pong handovers, or account for the unique characteristics of the new serving satellite~\cite{ahangarpour2024trajectory}. While our \textit{handover-aware} policy still outperforms the default WebRTC configuration, its performance is fundamentally constrained by the quality of these predictions. Therefore, developing a robust, fine-grained handover predictor is a critical next step.

\section{Conclusion}
\label{sec:conclusion}
Our work tackles key challenges in adapting existing RTC stacks to LEO networks. We propose a \textit{handover-aware} queue management policy that dynamically adapts the maximum pacing queue limit based on anticipated handover dynamics. Through a series of experiments, we
illustrate that our offline approach generalizes from offline telemetry logs to online deployments. We then provide initial results showing that in sparse LEO settings, our dynamic queue policy yields up to a $3$x improvement in video bitrates and reduces video freeze
rates by up to $62\%$ when compared to WebRTC's default queue policy in emulated sparse constellations. In contrast, our method generalizes beyond emulation and delivers up to a $41\%$ reduction in freeze rate and $40\%$ decrease in mean packet loss on real Starlink constellations compared to WebRTC's default queue policy.

\bibliographystyle{IEEEtran}
\bibliography{bibliography}

\end{document}